\documentclass{infor}
\volume{}
\issue{}
\pubyear{2025}
\articletype{research-article}
\doi{10.1515/9783111085722-019}

\theoremstyle{remark}

\makeatletter
\def\rtitleheadline{%
  \hfill {\small \thepage}\break
  \print@doi
}
\makeatother

\begin{document}
\begin{frontmatter}
\pretitle{Research Article}

\title{Big Data and the Computational Social Science of Entrepreneurship and Innovation}

\author[a,c]{\inits{Li.}\fnms{Ningzi} \snm{Li}\ead[label=e1]{ningzi@uchicago.edu}\bio{bio1}}
\author[b,c]{\inits{Lai.}\fnms{Shiyang} \snm{Lai}\ead[label=e2]{shiyanglai@uchicago.edu}\bio{bio2}}
\author[b,c,e]{\inits{Evans.}\fnms{James}
\snm{Evans}\thanksref{f2}\ead[label=e3]{jevans@uchicago.edu}\bio{bio3}}

\thankstext[type=corresp,id=f2]{Corresponding author.}
\address[a]{\institution{Booth School of Business, University of Chicago}, \cny{USA}}
\address[b]{\institution{Department of Sociology, University of Chicago}, \cny{USA}}
\address[c]{\institution{Knowledge Lab, University of Chicago}, \cny{USA}}
\address[e]{\institution{Santa Fe Institute}, \cny{USA}}

\begin{abstract}
As large-scale social data explode and machine-learning methods evolve,
scholars of entrepreneurship and innovation face new research opportunities but
also unique challenges. This chapter discusses the difficulties of leveraging large-scale
data to identify technological and commercial novelty, document new venture origins,
and forecast competition between new technologies and commercial forms. It suggests
how scholars can take advantage of new text, network, image, audio, and video
data in two distinct ways that advance innovation and entrepreneurship research.
First, machine-learning models, combined with large-scale data, enable the construction
of precision measurements that function as system-level observatories of innovation
and entrepreneurship across human societies. Second, new artificial intelligence
models fueled by big data generate ‘digital doubles’ of technology and business, forming
laboratories for virtual experimentation about innovation and entrepreneurship
processes and policies. The chapter argues for the advancement of theory development
and testing in entrepreneurship and innovation by coupling big data with big
models.
\end{abstract}

\begin{keywords}
\kwd{Entrepreneurship}
\kwd{venture funding}
\kwd{creative destruction}
\kwd{big data}
\kwd{digital doubles}
\kwd{embeddings}
\kwd{virtual experiment}
\kwd{artificial intelligence (AI)}
\kwd{large language models (LLMs)}
\kwd{deep neural networks (DNNs)}
\end{keywords}

\end{frontmatter}

\section{Introduction}

The availability of large-scale social and cultural data has erupted across the social sciences,
and has expanded the scope and precision of research in entrepreneurship and
innovation. Passive sensors, from social media to cell phone apps, yield rich traces of
human and entrepreneurial cognition, communication, and behavior \citep{AcevesEvans2023,evans2016machine}. Online platforms like Crunchbase and digital databases like
Pitchbook and VentureXpert have also made company, transaction, research, and invention
data and commentary more accessible than ever before \citep{guzman2023measuring,mcdonald2020parallel,zhang2020join}. Beyond observational data, modern
sensors have enabled the instrumentation of new venture experiments that enable
causal inference regarding the impact of organization, composition, and communication
on entrepreneurial outcomes \citep{hasan2019conversations,wright2023judging}. In parallel with the rise of big, unstructured social data, powerful machine learning methods have emerged that compress these data into operational measures
that can fit within and expand the scope of traditional statistical analyses. Even more
recent artificial intelligence (AI) models have empowered the creation of data-driven
simulations or digital doubles that facilitate in silico experimentation of social and organizational
life \citep{vicinanza2023deep}. These new models propose hypotheses
for empirical observation and experimentation \citep{AcevesEvans2023, evans2022text}.

Large-scale social and business data are relevant to understanding many aspects
of entrepreneurship and innovation, as they are also to many aspects of organizational
behavior and science and technology studies (STS). Nevertheless, innovation
and entrepreneurship face three special challenges that benefit more directly from
new data and new models than traditional organizations and STS research. First, successful
innovation and entrepreneurship are characterized by novelty. New ventures
involve novel combinations of technology, business opportunities, and skilled talent
to build new products and services that compete with potential rivals. When new
firms experience outsized success, it unleashes a process of creative destruction \citep{schumpeter2013capitalism}, whereby the novel organization clears the field and ‘destroys demand’
for competing technologies. Modeling entrepreneurial ventures and predicting
their success requires the identification of venture novelty, which imposes a unique
data burden on entrepreneurial science. Why? The amount of data required to identify
an organizational or market trend is at least twice that required to characterize a
modal tendency. The data required to identify whether a new venture represents a
novel combination of business or market elements, however, require a nearly complete
characterization of all that went before. In other words, novelty estimation requires
models sufficient to explain the vast majority of new and existing organizations,
and this, in turn, requires much more data and computation than identifying a
tendency or trend \citep{shi2023surprising, uzzi2013atypical, wang2017bias}.

A second, related special challenge concerns the study of origins and is shared by
scholars and scientists who focus on origins questions in other domains. Data are
forged, structured, and preserved through stable, mature systems. In the social sciences,
data are collected and preserved by institutions (e.g., from ancient temples to
modern states and corporations). In the natural world, data are preserved when conditions
are stable and persist over long periods of time. As a result, the origins of
space, time, matter, life, biological species, human language, cities, emerging organizations,
technologies, and ideas possess fewer data. By the same token, entrepreneurial
ventures are not required to financially report many details of their operations.
Unlike large, stable firms, new ventures rapidly evolve and pivot in relation to emerging
opportunities \citep{fink2019much, ries2011lean}. This instability and lack of reporting
requirements means that much less data are available on new ventures than persistent
firms, making systematic science more challenging. Similarly, innovations in
science and technology are least likely to be accounted for accurately in history \citep{merton1957priorities,stigler1980stigler}. Thomas Kuhn demonstrated that the most innovative discoveries
and inventions are typically attributed incorrectly, because they take time and
experimentation to understand (Kuhn, 1962). Fortunately, data are increasingly available
in the context of new venture emergence, ranging from the environments of
emerging firms to the digital traces of business discourse contemporary to them. New
ventures, like all new organizations, reflect their environments more than established
firms \citep{stinchcombe2013social} and so increased data, newly available for representation
and analysis, provide the entrepreneurial analyst with rich new opportunities for understanding.
Vera Rocha and Theodor Vladasel (Chapter 13, this volume) share a similar
concern in their chapter and they actively pursue new forms of data to tackle this
challenge.

A third special challenge involved in the study of entrepreneurship and innovation
relates to the difficulty associated with identifying functional equivalence among
new venture and innovation components. As described above, creative destruction involves
the replacement of one product, service, component, company, or industry
with another. This has historically been traced but not anticipated by entrepreneurship
scholars \citep{AkcigitVanReenen2023, aghion1990model}. Most commonly,
scholars measure functional equivalency ex post with the diffusion of a successful
innovation or the emergence of a successful firm \citep{Abbott1992,mizruchi1993cohesion,caballero2010creative,mizruchi1993cohesion}. Sometimes scholars have studied it negatively
through the failure of other products and firms `nearby' \citep{borchert2010creative,foster2010productivity}. Building a predictive science of innovation requires new data on the
character of products and services that would allow us to predict their functional relationships.
Fortunately, digital data are increasingly available on the description and
semantic content of products and services. Combined with dramatically improved AI
models, like neural network transformers that can discern subtle differences in meaning,
these data hold the potential to qualitatively anticipate and automatically compare
inventions, products, and companies at both high resolution and scale.

These three challenges suggest an incongruity. On the one hand, they suggest the
importance and potential of large-scale data for transforming our understanding of
entrepreneurship and innovation. On the other, they highlight the unique challenges
that entrepreneurial research has faced to assemble and benefit from those data. The
result has been that much entrepreneurship research has remained qualitative in
character. For this reason, much business school pedagogy on entrepreneurship has
been ceded to expert practitioners with a handful of mixed entrepreneurial experiences,
who are presumed to have insights of comparable or even superior legitimacy to
those of scholars who have studied a roughly equal handful of business cases. Pedagogy surrounding innovation is further stymied by the innovation paradox, where as successful principles of innovation become institutionalized, they cease to be a source of innovation in business and society \citep{cao2022destructive}.

In this chapter, we explore how demand for data in the study of entrepreneurship
and innovation is increasingly being met with new text data on entrepreneurial communications and descriptions, network data on social and institutional relationships,
and images, audio, and video data associated with new firm environments, inputs,
and outputs. We further explore how the recent shift from measurements to models
of large-scale data allow us to create digital doubles of public and private innovation
systems, which function both as system-level observatories for descriptive understanding,
on the one hand, and virtual laboratories enabling detailed simulation that
could accelerate theoretical development and testing, on the other. These properties
are especially important for entrepreneurship and innovation because data for observations
are sparse and formal experiments are difficult except in the earliest stages of
new venture creation \citep{koning2016field}. We argue that new data-driven models of
entrepreneurship and innovation hold the potential to raise the quality of these sciences,
transform education based on them to outcompete personal experience and anecdote,
and even contribute to the emerging trend of data-driven investment, selection,
and even technological and entrepreneurial catalysis.

\subsection{New Data Relevant to Entrepreneurship and Innovation}

The advancement of computational research within entrepreneurship and innovation
hinges on leveraging emerging opportunities presented by big data. New data opportunities
hinge on (1) the availability of new sources available for probing entrepreneurship,
innovation, and the business and technological background against
which they arise; and (2) new methodologies that can ‘datify’ unstructured information
into representations that can be tapped to yield new insights. In the paragraphs
that follow, we delineate emerging new data in four significant data categories. These
include network data, text data, image data, and audio data. Then we discuss two
broad approaches to these sources of information. The first is the data approach,
whereby information is extruded or distilled into rectangular matrix form and
treated like any other received or collected data for traditional statistical analysis and
causal inference regarding entrepreneurship and innovation. The second is the model
approach, which creates a data-driven digital double of the system from which innovation
and entrepreneurship emerge and uses it both to observe these phenomena
like a high-resolution telescope, and to generatively simulate these phenomena to facilitate
virtual experiments and hypothesis testing. We argue that data and modelbased
approaches are complementary, and together unleash the power of emerging
data and computation.

\subsubsection{Network Data}
Network data have long represented the vanguard of data-driven investigations in entrepreneurship
and innovation. Historically, researchers delved into a wide range of
relevant network forms. Traditional use of network data and analysis related to business
and new venture innovation includes interlocking directorates, which shed light
on corporate influence and diffusion through shared board memberships \citep{davis1991agents,heemskerk2013rise}; venture capital co-investment networks \citep{podolny1997better, podolny1999choosing,werth2013co}; patterns of employee migration between
firms, which illuminate the transmission of knowledge and the spatial dispersion
of skills \citep{woodruff2007migration}; hierarchical schematics (or org-charts) within organizations,
clarifying the distribution of power and pathways of information flow \citep{ennis1961men,Owenarticle}; input–output tables, which provide a macroeconomic
perspective on the transactions between firms and sectors \citep{mcnerney2009network};
collaboration networks that connect inventors and researchers, highlighting joint efforts
propelling innovation \citep{guimera2005team}; and networks defined
by interconnections among technology and knowledge themselves, analyzed
through patent classification to assess the emergence and spread of technological innovation
\citep{pennings1992technological, shi2015weaving,shi2023surprising,sourati2023accelerating}.
Each of these data forms has contributed to our understanding of mechanisms underpinning
entrepreneurship and the genesis of innovation.

\subsubsection{Text Data}
Text data have become the most plentiful form of data available for analysis about entrepreneurship and innovation, given their critical role in signaling novel value and catalyzing potential exchanges \citep{evans2016machine}. In the traditional entrepreneurship and innovation research, text analysis was either qualitative and interpretive in nature, or reduced text to tabulations of word counts. Company materials,
investor communication, organizational histories, and founder interviews provided narrative depth but lacked scalable analysis \citep{giudici2005applied}, except using brittle dictionary methods that lacked a holistic understanding of meaning in context \citep{roundy2019understanding,suarez2021entrepreneurship}. Patent documents, often parsed for their classification codes, offered glimpses into technological progression \citep{fleming2001technology,thompson2005patent}.

Newly accessible data streams have enriched the textual landscape. (1) Rich descriptions of firms, as narrated in the fast-paced arena of business news and online information services (e.g., Crunchbase, PitchBook, VentureXpert), offer a granular view of company evolution designed to serve active investors and job seekers \citep{guzman2023measuring}. (2) The textual content of proposals reveals intentions and strategies
underlying new business endeavors \citep{bromham2016interdisciplinary}. (3) Patents and academic publications trace the trajectory of discovery and technological advancement from conception to implementation \citep{bromham2016interdisciplinary,hofstra2020diversity,park2023papers}. (4) Product fact sheets and descriptions provide detailed accounts of product innovations,
embedding technical specifications within the language of market appeal and consumer need \citep{silver2022balancing}. (5) Employee reviews and company responses on platforms like GlassDoor furnish insights into company culture and employee experience, providing a dual perspective of firms as workplaces and self-conscious business entities \citep{campbell2022tone,dube2021disciplinary}. (6) Business-oriented social media profiles, particularly on platforms like LinkedIn, offer a crowd-sourced wealth of data on professional networks, skill distributions, and industry trends \citep{Alaql2023}. (7) Emerging access to internal corporate communications, such as emails and instant messages, have allowed researchers to observe the informal and formal discourse within companies \citep{zha2016unfolding}. In
all, tapping these rich veins of newly available text, researchers can paint a more comprehensive and nuanced picture of the entrepreneurial and innovation landscape, providing insights broader in scope and deeper in detail than before.

In the `text as data' approach, people engage with the wealth of textual information by converting the unstructured raw text into structured, analyzable forms \citep{evans2022text,gentzkow2019text,grimmer2022text}. This process often involves text-mining techniques that transform narratives and qualitative content into quantifiable features. Through this lens, a patent transforms into a
broad array of indicators reflecting innovation, similar to how Gianluca Carnabuci and Balázs Kovács argue in their chapter on patent data (Chapter 14, this volume) that these data will complement traditional patent variables. GlassDoor reviews are distilled into metrics of employee sentiment, and the sprawling narratives of firm histories are encoded into timelines and trends. These preprocessed, structured, and
spreadsheet-friendly data allow researchers to apply statistical models to text, creating an empirical basis for measuring variables such as sentiment and topical prevalence. Variables can be constructed to represent abstract concepts like innovation, strategic orientation, and distinctiveness, making it possible to conduct large-scale studies that correlate these aspects with business outcomes \citep{bellstam2021text,guzman2023measuring,taeuscher2022categories}. Another approach draws upon data compression techniques, from matrix factorization \citep{dumais2004latent} to topic modeling \citep{chandra2016mining,singh2023evolving} to neural auto-encoding \citep{AcevesEvans2023,DBLP:journals/corr/abs-1905-12741}, to generate unnamed features that distinguish documents and statements in ways that reflect expressed meanings. By quantifying the qualitative, this approach enables a new dimension of analysis, turning textual artifacts into rich sources of data ripe for hypothesis testing and causal inference.

The `text as model' approach enables us to delve into a more sophisticated use of textual data that transcends traditional analysis and draws on the logic of Alan Turing's `Imitation Game' \citep{turing2009computing}. Deep-learning algorithms act as the modern engine for this approach, enabling us to encode vast amounts of unstructured text into a model space that retains a compressed description of the data, while enabling the simulation of counterfactual text. These models may encapsulate the essence of individual firm communications, while simultaneously capturing the nuanced tapestry of social, cultural, and strategic contexts within which new enterprises and technological breakthroughs take shape. By transforming raw text into multi-dimensional representations\footnote{These representations are considered high dimensional with respect to traditional two or three dimensional models of meaning common in cognitive and cultural science \citep{osgood1957measurement}, but low dimensional with respect to the vast number of distinct words used in discourse, which might otherwise be each represented as a categorically independent dimension.} with minimal distortion, the models do more than just provide retrospective
insight. They can serve as virtual laboratories in which we can simulate and predict the future \citep{evans2022text}. We can forecast the ripple effects of innovation, predict market responses to new products, and understand how shifts in cultural and economic climates might influence entrepreneurial success or failure \citep{AcevesEvans2023,foster2015tradition}. Furthermore, this approach allows us to explore the interconnectedness of elements within the entrepreneurial ecosystem. For example, through embedding texts, one can elucidate how the interplay between emerging technologies and regulatory frameworks might shape the viability and adoption of innovations \citep{belikov2020detecting}.

\subsubsection{Image Data}
Image data also play an emerging role in the landscape of entrepreneurship and innovation research. While traditional approaches have relied on qualitative analyses of corporate logos and visual intelligence gathered from firm infrastructure, settings, and interpersonal interactions \citep{dowling2000creating}, the advent of new data forms is adding a powerful new dimension to our understanding of corporate identity and
culture. 

Contemporary datasets are beginning to encompass visual content that could provide deeper insights into corporate and product branding. This visual repository includes: (1) video footage which spans slick advertisements designed to entice consumers, to company profile videos that communicate corporate values and strategic direction \citep{li2021product}. (2) The curated images of founders and key personnel are more
than mere headshots \citep{choudhury2019machine}. In the same way that mugshots have begun to revolutionize the analysis of judicial decisions \citep{ludwig2022algorithmic}, corporate headshots reflect a company's diversity and ethos, and can even become a part of the company's innovation story, resonating or repelling potential customers and investors \citep{kamiya2019face}. (3) Product photographs, beyond showcasing features and design, can subtly communicate signals regarding innovation and quality associated with a brand, forming an essential part of a firm’s visual lexicon \citep{bu2022multilevel}. Moreover, (4) personal and corporate profiles on business social media, like LinkedIn, offer a gallery of personas and impressions that convey public identities \citep{nguyen2021towards}. Images shared across social media platforms extend this narrative by portraying company culture, events, and the day-to-day reality of work environments, which can be mined with image analytics to characterize otherwise invisible distinctions and similarities.

The `image as data' approach aims to systematically decipher visual data points. By extracting sentiment and predicting attributes such as identity (e.g., demographic inferences like the prevalence or sparsity of ubiquitous ‘white men’) or diversity levels, researchers can incorporate these insights into traditional statistical models. As with text, image2vec approaches can encode images, generating coordinates in a high dimensional image space that characterize meaningful differences in the objects and settings represented within them \citep{jo2018development}. An image is worth a thousand words, and many of these models contain nearly a thousand dimensions that allow complex proximity assessments. These methods enable the quantitative analysis of visual elements, providing a structured way to evaluate visual branding, firm composition, and cultural dimensions of its experienced environment.

Pushing the envelope, the ‘image as model’ approach takes a different approach to visual data analysis. Encoding images through advanced computational techniques like generative adversarial networks (GANs) \citep{ludwig2022algorithmic} has enabled the discovery of new dimensions that define corporate and product imagery \citep{hua2007discriminant}. Generative image models also facilitate the automatic production of new images and video footage, thereby enabling a predictive understanding of visual trends and their implications for innovation and brand perception \citep{li2021product}. This modeling technique has the potential to unveil patterns and insights that can inform strategic branding decisions and forecast emerging visual trends that characterize the entrepreneurial and innovation ecosystem.

\subsubsection{Audio Data}
Audio data, compared with the previous three forms of data, are markedly less explored in the context of entrepreneurship and innovation research. Researchers classically collected audio data primarily through interviews and communication events, such as press conferences and investor pitches. These auditory snapshots, though limited, were reduced to text and provided critical insights into the rhetoric and narratives that business leaders used to influence stakeholders and shape public perception \citep{mauney2004creating,pennebaker2002language}.

In the current digital era, the landscape of audio data in entrepreneurship and innovation research has expanded significantly, encompassing a variety of new forms
that enrich the auditory dimension of business analysis. (1) Advertisements, including not only text but also audio, for example, can be mined to induce firms’ strategic communication and positioning \citep{rodero2020your}. (2) The prevalence of online meeting applications, like Zoom, have generated an expanding array of audio data from business
meetings, ranging from virtual new venture pitches to stakeholder meetings \citep{chakraborty2024maud}. (3) Furthermore, social media has
arisen as a fertile ground for entrepreneurial voices, with audio broadcasts emerging as vital channels for the dissemination of entrepreneurial ideas and insights \citep{grewal2021marketing}. (4) The rich audio tapestry of investor pitch sessions, particularly those broadcast on entrepreneurial television programs, provides unique opportunities to dissect the communication strategies that correlate with fundraising success \citep{markowitz2023authentic}. Finally, (5) video that synchronizes audio with images is increasingly available for analysis using new tools for video learning, modeling, and understanding \citep{kaminski2017user}.

The `audio as data' approach focuses on the extraction and quantitative analysis of information from audio recordings. Through advanced signal processing and natural language processing techniques, researchers can convert speech into accurate transcripts, capturing the textual content of verbal communications. Transcription facilitates the analysis of language use, communication style, and information exchange. Beyond mere transcription, this approach extends to the assessment of
speech sentiment, the stance of the speech, and speaker identity recognition \citep{fan2010speaker,rao2021sentiment}. In sum, this approach harnesses the power of computational analysis to distill audio waveform data into variables, which can be analyzed with statistical models to uncover patterns and draw conclusions
about entrepreneurial behaviors, strategies, and outcomes.

The `audio as model' approach encodes the audio data into deep neural networks (DNNs) to preserve rich and detailed information and enable simulation. This approach allows analysts to identify latent dimensions within audio. Researchers have extracted tagging features through optimized DNN models to facilitate the building of
a virtual business assistant for audio tagging tasks \citep{el2023optimized}. Alternatively, it offers the potential to simulate future audio scenarios based on identified patterns \citep{beguvs2021ciwgan}. By identifying the acoustic signatures of successful business communication, it becomes possible to generate simulated voices and audio environments \citep{purdy2023sound}. The synthetic audio embeddings that result could serve as an immersive green field for investigation, supporting experimental platforms that allow researchers and industry actors to investigate effective communication and presentation techniques for showcasing business innovation. By mirroring situations that have historically resonated with investors and stakeholders, these environments facilitate the exploration of strategies that lead to successful engagements. As with image and video models, audio modeling raises complex ethical questions. Insofar as entrepreneurs and innovators can learn how to convey their ideas more persuasively, investors and decision makers will need better discriminative models in the arms race most visible in the generation and filtration of mis- and disinformation in society.

While effective utilization of new data across the four forms described above unveils fresh avenues for investigating entrepreneurship and innovation, their integration augments this potential by enabling construction of a data-driven digital double of the entrepreneurial or innovation target under analysis. For example, combining
image and text data from business social media, analysts can more closely measure and model how entrepreneurs and investors present themselves to their human audiences \citep{martinec2005system}. Moreover, by interlinking different forms of data through simulated social, experiential, or cognitive processes, our predictions can
markedly improve. Consider one of our recent papers in which we sought to predict the distribution of new materials discovered to have valuable energy-related or therapeutic properties. To begin, we replicated predictions by modeling scientific text alone \citep{tshitoyan2019unsupervised}. We then interlinked properties and materials from article text with article authors by simulating the socio-cognitive process of discovery using random walks over the hypergraph of articles. For example, we began with a property (e.g., COVID treatment), then jumped to a random article with the property (e.g., `Effect of interferon alpha and cyclosporine treatment...on Middle East Respiratory Syndrome Coronavirus...') to a random author on the article (e.g., John Nicholls) to another random article by the author (e.g., `Evaluation of the human adaptation of influenza A/H7N9...') to a random author on the article (e.g., Michael Chan) to another random article by the author (e.g., ``Production of amphiregulin and recovery from influenza...'') to a random author on that article (e.g., Sabra Klein) to another article by that author (e.g., ``Progesterone-based therapy protects against influenza...'') to a random material on that article (e.g., Progesterone). These walks simulated the cognitive availability of a hypothesis that some material had a valuable property (e.g., Progesterone is an effective COVID treatment for men) through human experience and communication (e.g., Sabra Klein, who knew progesterone, had a conversation with Michael Chan, who conferenced with John Nicholls, who understood COVID), which led to a clinical trial and demonstration (e.g., at Cedar-Sinai Hospital connected with the University of California, Los Angeles.) Modeling this socio-cognitive process with millions of random walks instead of the text alone improved our predictions of discovered innovations by 400 percent for the case of COVID therapeutics, and an average of 100 percent across hundreds of diseases and electrical materials \citep{sourati2023accelerating}. Linking data through simulated processes of discovery, invention, and diffusion enables the identification of high-dimensional proximities that promote better measurement and modeling, as illustrated above. In this way, new data forms can be extruded into traditional variables for statistical analysis (e.g., text, images, networks `as data'), but the modeling approach offers richer, unexplored opportunities for research on entrepreneurship, innovation, and social and strategic dynamics more broadly \citep{AcevesEvans2023}.

\subsection{Big Data and Digital Doubles of Entrepreneurship and Innovation}
`Big' unstructured or semi-structured data is difficult to incorporate directly into social
scientific analyses of entrepreneurship and innovation. For example, the character
of constructed network data on the configuration of innovative ideas and technologies
often violates simple network analysis and metrics. When using context to
create the path distance between components, like patent classes involved in an invention,
most components within the system may be within two to three steps, the
`friend of a friend' \citep{shi2015weaving}. The resulting `hairball' networks are far too dense
to perform any traditional network analysis procedures without thresholding (e.g.,
stronger ties equal 1; weaker ties equal 0). Such choices remove important data and
miss critical phenomena, however, like Granovetter's well-known `strength of weak
ties' within a system \citep{granovetter1973strength}. More pervasively, big unstructured data
often can only become structured through inferring connectedness through proximity,
as with language models where words become linked through shared context
(e.g., Word2Vec) \citep{mikolov2013distributed}. These implicit
connections, crucial for finding patterns in and extracting structure from big
data, also represent hidden confounders that violate traditional statistical models'
standard independence assumptions. The pervasive presence of `social influence' in
big data creates a web of interdependencies among data points that starkly contrasts
with the small-scale, sparsely interconnected datasets previously utilized by researchers
that allowed analysts to preserve the illusion that they satisfied restrictive statistical
assumptions \citep{karlsson2023detecting}. This has also led complex correlations in
big data to become misconstrued as causal links.

Another challenge with big data is that the compression of multi-dimensional
data into fewer predictive variables can lead to significant information loss. This dimension
reduction process tends to favor confirmation over the discovery of novel insights,
as it overlooks the capacity of big data to illuminate unanticipated patterns and relationships. In image recognition, dimensionality reduction techniques that
simplify data into principal components may overlook subtle visual cues critical for
identifying emerging patterns. For example, studies in facial recognition have demonstrated
that image compression algorithms such as `Eigenface' adversely impact system
accuracy when encountering novel facial expressions or features not well represented
in the training set \citep{mulyono2019performance}. Consider a recent policy paper that sought to predict whether judges
would grant bail. Computational economists Jens Ludwig and Sendhil Mullainathan
built a GAN model to encode the booking photos of the accused, then gave the face
model and unnamed dimensions most predictive of positive bail determinations to
human annotators \citep{ludwig2022algorithmic}. The annotators could intuitively
recognize these previously unnamed dimensions as grooming and heavy-facedness.
Better-groomed and heavier-faced defendants were much more likely to be granted
bail and set free. These discovered qualities individually explained almost as much in
predicting judgments as gender (i.e., women were also much more likely to be set
free). If the authors had instead projected the data down to theorized dimensions,
they would not only have lost the opportunity to discover new forms of bias in the
system, but also have strongly sacrificed predictive capacity.

By accounting for complex variable interactions, data-driven models can resolve
these challenges and better predict and simulate complex outcomes such as those underlying
entrepreneurial and innovation success. We highlight the particularly promising
strategy of developing data-driven models that create digital doubles of entrepreneurial
firms, innovative systems, and their underlying processes including
strategic search, decision making, knowledge dissemination, quality assurance, and
complex competition \citep{domingos2018our}.

Digital doubles act as data-driven virtual counterparts to their real-world entities.
They are increasingly used in domains where first-principles models are insufficient
to explain and predict the majority of variation in a phenomenon. Even in the most
‘organized’ natural scientific subjects, such as chemistry, satisfiable prediction
through first principles alone is unattainable. Simple models of bond formation do
not equip us to forecast either the structure or function of complex structures like
polymers or proteins. In protein folding, the AlphaFold models, developed by Google
subsidiary DeepMind, facilitate protein structure prediction that so dramatically outstrip
first-principles models with a digital double based inferred on a large language
model (LLM) architecture \citep{senior2020improved} that it received the 2024
Nobel Prize in Chemistry. This gap between first principles and outcomes is particularly
pronounced in the complex and dynamic arenas of entrepreneurship and innovation,
where the interplay of myriad factors leads to unforeseen discoveries and
technology breakthroughs \citep{cao2022destructive}. In these domains, we argue that the role
of digital doubles could be invaluable, offering rich, predictive simulations to navigate
the complex and evolving landscape where traditional models cannot.

Data-driven digital doubles are characterized by a twin character, as system-level
observatories and virtual laboratories. They represent system-level observatories in the
form of high-dimensional ‘embeddings’ that replicate simulated systems with fidelity,
allowing for data analysis that mirrors system dynamics with minimal distortion. These
embeddings open new opportunities for analyzing entrepreneurship and innovation.
For example, in measuring technological similarity, past researchers have typically depended
on a code-based methodology, utilizing classifications such as the American Patent
Classification, and International or Cooperative Patent Classification (IPC/CPC). By
projecting patent texts into an embedding space constructed from text and diagrams,
however, researchers can capture a broader spectrum of patented technologies \citep{AcevesEvans2023,li2018deeppatent,shi2023surprising}. Technology terms, patent classes,
patented technologies, and corporate patent portfolios can all be projected to these
spaces, with higher-level groupings (e.g., patent portfolio) the centroid, or high-dimensional
average, of lower-level groupings (e.g., technology patent class). This approach
allows for the precise calculation of similarities between technology units as cosine
distances within an embedding space \citep{hain2022text}.
Furthermore, they allow us to infer innovation-relevant dimensions within these
embedding spaces, such as how technologies array along the `medical' vs. `engineering'
dimension, which may be calculated simply as the cosine distance of a patent and the
subtraction of the word vectors for `medical' and `engineering' \citep{kwak2020frameaxis, An2018SemAxis,kozlowski2019geometry}. Embeddings also enable the identification
of self-labeling coordinates around which technologies cluster, which represent a more
precise and scale-able advance in context-sensitive topic modeling \citep{ArsenievKoehler2022}.

Insofar as `digital doubles' are constructed as generative models (e.g., LLMs, described
in detail below), they can also function as virtual laboratories, enabling researchers
to conduct simulations of various innovation scenarios, such as the merger
of different technologies, the implications of a technology applied to a new market
application, or the receptivity of an audience to a new product, service, or firm. This
is a complementary approach to the field experiments discussed by Chiara Spina and
Sharique Hasan (Chapter 20 of this volume) to better understand idea formation in
innovation and entrepreneurship. Simulations provided by digital doubles could extend
to the realm of policy, where digital doubles can test the impact of policy shifts
on innovation and entrepreneurial activity \citep{kumar2021digital}. By harnessing this dual capability, digital doubles offer an expansive toolkit for
researchers and policymakers alike, facilitating a deeper understanding of complex
systems and the exploration of future possibilities.

\subsection{Digital Doubles from Deep Neural Network Transformers}
DNNs are currently a dominant architecture for digital doubles \citep{domingos2018our}. DNNs are a class of data-driven models, characterized by an ensembled architecture
of many nonlinear models, roughly analogous to neural synapses in the brain,
interconnected across layers from data inputs to predicted or generated outputs. The
outcomes of lower-level models feed inputs to higher-level models that ultimately predict
desired outcomes or generate desired outputs. Distinguished from statistical and
even other machine-learning methods, DNNs are designed to search through vast
spaces of interactions between variables and identify those most predictive of the outcome
in question. This exploration is not limited to merely processing observable inputs;
rather, DNNs synthesize novel, synthetic variables within their enigmatic internal
constructs known as `hidden variables.' These variables, which are not explicitly
part of the input data, emerge through the network's learning process, reflecting complex,
nonlinear relationships that the network infers from the data itself. `Learning' in
a DNN typically relies on an optimization procedure involving the sequential updating
of nonlinear regression weights using the partial derivative of the model's error,
propagated through the chain rule of calculus, to a proposed adjustment to each
weight in a process called back-propagation \citep{miikkulainen2024evolving}. In this way, a critical DNN capacity involves their capacity to model, predict,
and discover mechanisms underlying complex systems. Many DNNs are `feed forward'
models, tuned to predict targeted outcomes like complex nonlinear regressions.
These might be trained to anticipate when and whether an entrepreneurial firm
might go public, or if a patented technology will become a `hit.' Other DNNs are `autoencoders,'
tuned to encode, predict, and `describe' their input data, like complex nonlinear
factorization models (e.g., PCA, SVD, factor analysis). These might create a vector
space that compresses evolving business discourse \citep{cao2023higher} and
allows one to calculate the distance between firms within that space \citep{AcevesEvans2023}.

Transformers represent a new family of DNN generative auto-encoders introduced
by Google researchers in 2017 \citep{vaswani2017attention}, which
underlie modern LLMs. LLMs encode language sequences into powerful models that
form digital doubles of the discursive systems they represent, just as the AlphaFold
transformer forms a digital double of the proteins it simulates. These models include
multiple layers of `self-attention,' where words or other system components are predictively
linked to one another in a complex matrix that maximizes their ultimate
ability to predict masked or future words. Through this training, ChatGPT and similar
LLMs become digital doubles of their training data (e.g., text from the internet) that
can model and reproduce what humans would write given prompts. LLMs do not
memorize exact utterances from their training texts but learn the latent functions capable of generating language conforming to shared discourse. An analyst can program
these models with natural language, prompting them with text to evoke a specific discourse.
The LLM then optimizes its response based on the underlying embedding that
captures syntactic, semantic, and even pragmatic relevance \citep{dai2022can,von2023transformers}. The newest LLMs are multimodal
models that not only encode text, but also images, audio, and other associated
data.

In the context of entrepreneurship and innovation, the power of transformers
can be used in the dual ways described above. On the one hand, transformers enable
the formation of high-dimensional, geometrically interpretable knowledge spaces
available for detailed analysis that function as observatories of the systems they encode.
Through these powerful macroscopes, entities (e.g., innovators, technologies,
new ventures, inventive regions) are rendered in context to reflect precise distances
between them based on complex linkages in the underlying data. Second, transformers
are generative models, and like ChatGPT, can function as virtual laboratories to
simulate outcomes of their represented system to experimental provocation (e.g.,
technical advances predicted to result from specific funding or within specific regions).
Transformers can also support the estimation of structural models to causally
identify the impact of funding on entrepreneurial success, regional proximity on collaborative
innovation, and a host of other potential models.

At the beginning of this chapter, we highlighted three unique challenges in the
study of entrepreneurship and innovation that could benefit from big data: the vexing
measurement of novelty in new ventures and innovations, the data-poor study of origins
in business and technology, and the need to identify functional equivalence in
order to measure processes of creative destruction. Here we illustrate how each of
these challenges can be supported by LLMs, and how each case highlights different
aspects of LLMs that can support entrepreneurial and innovation research.

LLMs support the evaluation of novelty by encoding the discourse of a society or
market within a comprehensive embedding space in which every entrepreneurial or
innovation-relevant concept, technology, use case, or new venture description can be
projected with precision. Insofar as the components of a technology, its linkage to a
business use case, or the language of a company description are highly probable outcomes
of the LLM, then they are not novel but rather follow the flow of discourse in
business and society. Insofar as these are highly improbable or perplexing to the
model, however, then they combine concepts, technologies, business cases, and market
scenarios in ways that are novel and surprising to the system. This usage of the
LLM underscores its capacity to encode the nature of the system such that technology
or new venture description, projected to its embedding, enables the model to simulate
the system’s surprise. This can be calculated not only with model perplexity, but also
when large distances are required to connect the innovation or new venture description
within the model’s high-dimensional embedding space. This allows for a sophisticated understanding of how disparate ideas or elements might coalesce to form innovative
products, services, and enterprises \citep{AcevesEvans2023}.

The second challenge in entrepreneurship and innovation research concerns the
`small data' available for the study of origins. New data streams have significantly expanded
the reservoir of relevant data, but for novel enterprises and inventions, detailed
information remains scant. This presents a formidable obstacle for those adhering
to the text/images/networks-as-data approach. In contrast, LLMs offer a powerful
resolution. Once LLMs are pre-trained on wide-ranging contextual data, such as seemingly
unrelated text from Wikipedia and the web, they can infer accurate relationships
from even scant facets of a new venture or technology. For instance, by processing
a corpus of historical business press releases, transformers could infer latent
connections and precise distances between those companies with no textual overlap
in their descriptions, drawing on voluminous contemporary business discourse from
newspapers, magazines, and the web. This highlights the ‘foundation model' aspects
of LLMs, where when built from vast stores of available language, they can effectively
expand sparse data. In the context of a world of discourse, even description data from
Crunchbase on three new ventures alone is enough to precisely identify the relative
distance or proximity between them and within the full matrix of societal concepts
and cultural dimensions.

\begin{figure}[htbp]
    \centering
    \includegraphics[width=0.8\textwidth]{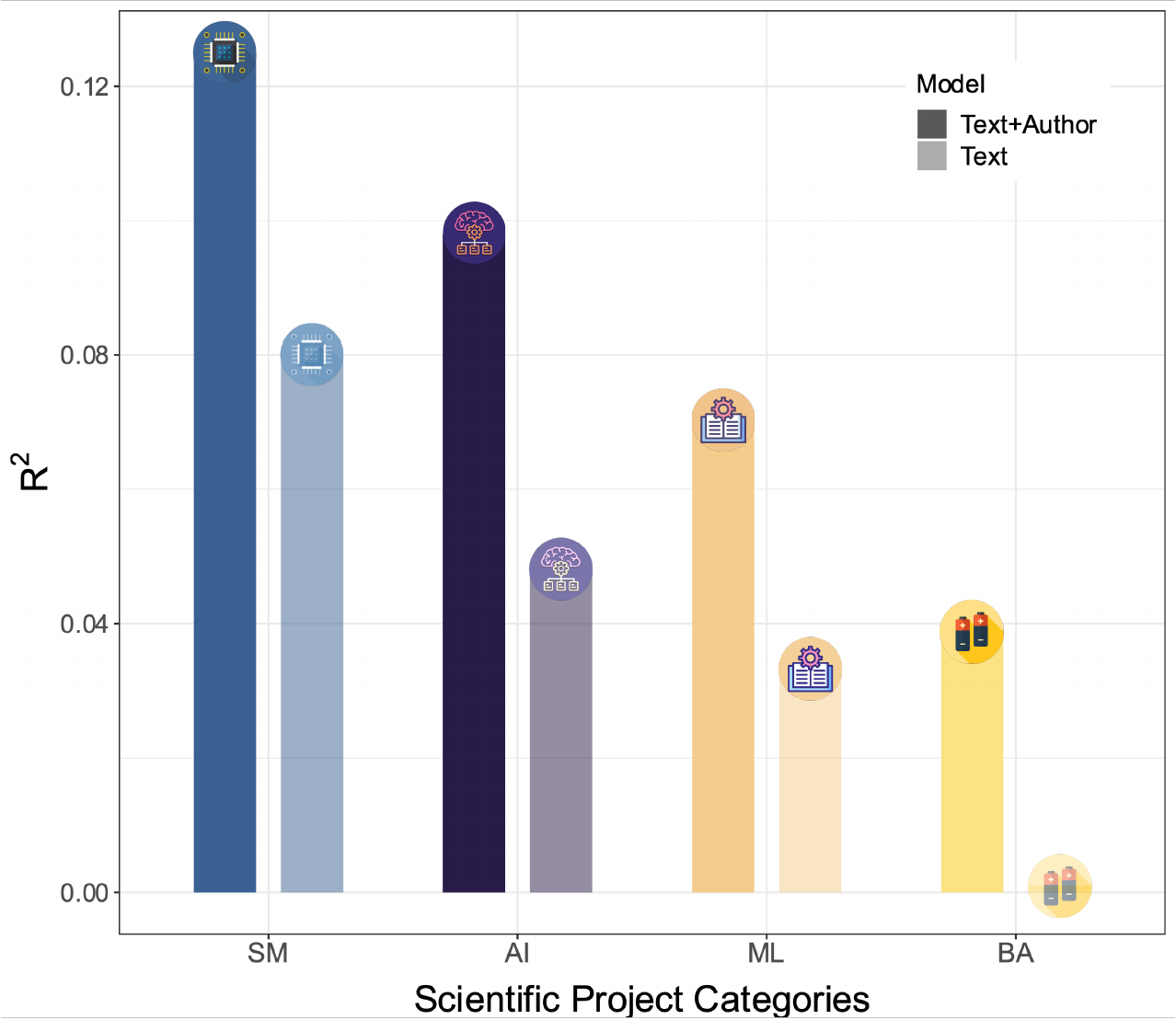}  
    \caption{The comparative r-squared values for linear model predictions of performance across four categories of scientific projects (performance is measured by parameter number for ML/AI models [ML]/ [AI], current density for semiconductors [SM], energy density for batteries [BA]; the two bars for each category contrast the predictive power of models using only semantic embeddings against those enhanced with both semantic and author embedding information.}
    \label{fig:your_label}
\end{figure}

The third challenge in entrepreneurship and innovation that can be approached
with big data relates to complex synergies between technologies and their functions
in markets and society. Without being able to anticipate the functional overlap or
equivalence of technologies for specific business markets, then the central process of entrepreneurship and innovation—creative destruction, whereby improved technologies and businesses make their predecessors obsolete—cannot be predicted or positively identified. LLMs can be made much more accurate for context-sensitive measurements through two related processes. First, they can be `pre-trained' with text data of special relevance to the context in question. In order to predict the functional
equivalent of a technology, LLMs can be pre-trained on technology patents following
their initial pre-training on large-scale text (e.g., a sample of the web). This might profitably include the `claims' section in which the patent’s inventor articulates the technology's
legally protected uses. Furthermore, LLMs can also be `fine tuned' to predict
other data that complement and improve their performance on a focal task. In a
small-scale analysis, we found that integrating author collaboration networks with
scientific textual embeddings substantially improved accuracy in predicting the technology performance compared with text alone, as shown in figure \ref{fig:your_label}. Yong-Yeol
Ahn and colleagues \citep{lee2024neural} fine-tuned a sentenceBERT
LLM, which markedly improved their prediction of new journals and new collaborations.
Fine-tuning transformer models with data from complementary systems can
yield superior results that enable the kind of precise functional equivalence required
for anticipating technology replacement.

\subsection{Big Data and the New Science of Entrepreneurship and Innovation}

The business processes of innovation and entrepreneurship are deeply entwined. On
the one hand, new ventures represent a special instance of socio-technical innovation. On the other, entrepreneurial firms catalyze social and technological innovations that serve societal functions. In this chapter, we have reviewed novel data streams relevant to entrepreneurship and innovation in the form of networks, text, images, and audio that could enable us to deepen our understanding in these fields. We have also reviewed two broad approaches through which they can be harnessed for analysis: as data and as models. For researchers reliant on traditional data forms and conventional linear or nonlinear models, the influx of new data offers the chance to substantially increase sample sizes and feature richness, improving both description and prediction. Despite this advantage, it does not fully leverage the intrinsic value of these new data. The traditional data approach compresses the granularity of semi- or unstructured
data, reducing its power to facilitate novel discovery or improve prediction.

Here we seek to demonstrate the comparable power of the model-based approach
to big data as a superior approach method for many potential analyses in the context of entrepreneurship and innovation research. This approach may leverage the power of
DNNs and contemporary transformers in order to intuitively encode complex, unstructured data into high-dimensional embeddings, bypassing the need for predefined features
and low-dimensional representations. By building data-driven digital doubles of
entrepreneurial and innovative phenomena, researchers can construct innovation observatories to survey novel technologies and new firms, yielding instantaneous, actionable insights. Furthermore, digital doubles can also function as virtual labs, permitting simulation system responses to new technologies and business ventures, explorations that might be untenable in the wild. Specifically, for the study of entrepreneurship and innovation, generative models can enable the ‘hallucination’ or envisioning of new technologies, the birth of novel firms, and the consequence of innovation investments. This forward-thinking strategy does not simply chart the trajectory of what has been; but forges counterfactuals for causal analysis, and paves the way for both a new science of entrepreneurship and innovation, and a method to accelerate successful investment,
venturing, and technology development \citep{rzhetsky2015choosing}.

\newpage
\bibliographystyle{infor}
\bibliography{biblio.bib}
\vfill

\end{document}